%% file: main.tex
\documentclass[a4paper,twoside]{article}

\usepackage{epsfig}
\usepackage{subcaption}
\usepackage{calc}
\usepackage{amssymb}
\usepackage{amstext}
\usepackage{amsmath}
\usepackage{amsthm}
\usepackage{multicol}
\usepackage{pslatex}
\usepackage{apalike}
\usepackage[bottom]{footmisc}

\usepackage{multirow}
\usepackage{algorithm}
\PassOptionsToPackage{noend}{algpseudocode}
\usepackage{algpseudocode}
\usepackage{booktabs}
\usepackage{hyperref}
\usepackage{comment}
\usepackage{circuitikz}
\usepackage{caption}
\usepackage{subcaption}

\usepackage{SCITEPRESS}     

\begin{document}

\title{CRGC  \subtitle{A Practical Framework for Constructing Reusable Garbled Circuits}}

\author{\authorname{Christopher Harth-Kitzerow\sup{1}*, Georg Carle\sup{1}, Fan Fei\sup{2}, Andre Luckow\sup{3} and Johannes Klepsch\sup{3}}
\affiliation{\sup{1}Technical University of Munich, \sup{2}Leibniz University Hannover, \sup{3}BMW Group}
\email{\href{mailto:christopher.harth-kitzerow@tum.de}{*christopher.harth-kitzerow@tum.de}}
}

\keywords{Secure Multiparty Computation, Garbled Circuits, Privacy Enhancing Technologies.}

\abstract{In this work, we introduce two schemes to construct reusable garbled circuits (RGCs) in the semi-honest setting. Our completely reusable garbled circuit (CRGC) scheme allows the generator (party $A$) to construct and send an obfuscated boolean circuit along with an encoded input to the evaluator (party $B$). In contrast to Yao's Garbled Circuit protocol, $B$ can securely evaluate the same CRGC with an arbitrary number of inputs. As a tradeoff, CRGCs predictably leak some input bits of $A$ to $B$.
We also propose a partially reusable garbled circuit (PRGC) scheme that divides a circuit into reusable and non-reusable sections. PRGCs do not leak input bits of $A$. 
We benchmark our CRGC implementation against the state-of-the-art garbled circuit libraries EMP SH2PC and TinyGarble2. Using our framework, evaluating a CRGC is up to twenty times faster, albeit with weaker privacy guarantees, than evaluating an equivalent garbled circuit constructed by the two existing libraries. Our open-source library can convert any C++ function to a CRGC at approx. 80 million gates per second and repeatedly evaluate a CRGC at approx. 350 million gates per second. Additionally, a compressed CRGC is approx. 75\% smaller in file size than the unobfuscated boolean circuit.}

\onecolumn \maketitle \normalsize \setcounter{footnote}{0} \vfill

\input{01_Introduction}

\input{02_OurApproach}

\input{03_Benchmarks}

\input{04_RelatedWork}

\input{05_Conclusion}

\bibliographystyle{apalike}
{\small
\bibliography{refs}}

\input{Appendix}

\end{document}

%% file: 01_Introduction.tex
\section{\uppercase{Introduction}}

Secure Multiparty Computation enables parties to execute functions on obliviously shared inputs without revealing them \cite{lindell2020secure}. Yao's Garbled Circuits protocol \cite{YAO1,YAO2} is a popular Secure Multiparty Computation protocol for realizing semi-honest two-party computation. Following the protocol, a circuit generator $A$ sends its encoded inputs and the encrypted and permuted gate output tables of a boolean circuit to a circuit evaluator $B$. $B$ can obtain only one encoded input per circuit through Oblivious Transfer. Thus, each time $B$ wants to obtain an output from a different input, it needs to request another garbled circuit. Garbled circuits get large in file size. Our Reusable Garbled Circuit (RGC) schemes allow $B$ to re-use a garbled circuit for multiple evaluations with different evaluator inputs. They significantly reduce communication overhead compared to sending a new garbled circuit for each evaluation. RGCs enable party $A$ to send obfuscated data to an untrusted party $B$ while ensuring that sent data remains secret and can only be used for its intended purpose implemented by the circuit. $B$ can evaluate an RGC with an arbitrary number of inputs without revealing $A$'s input. 

Existing RGC schemes usually rely on cryptographic primitives that are too complex for real-world use cases. Our key idea instead is to utilize information-theoretic techniques to obfuscate the wire labels that $A$ sends to $B$ in a way that hinders $B$ from learning $A$'s secret inputs. With this approach, $B$ can repeatedly evaluate the same obfuscated circuit with arbitrary inputs. Only when $A's$ input changes it needs to construct a new RGC. While constructing an RGC can take longer than constructing a garbled circuit, it pays off over time due to faster evaluation speed. Not all gates in a circuit can be obfuscated without leaking input bits. Thus, $A$ has two options: 

\begin{enumerate}
    \item It obfuscates only those gates with our techniques that do not leak information. It then groups the remaining unobfuscated gates into $n$ non-reusable sub-circuits and prepares $n$ Yao's Garbled Circuit protocols. With this approach, we obtain reusable and non-reusable sections in a circuit. We call the resulting circuit a partially reusable garbled circuit (PRGC).
    \item It obfuscates all gates with our techniques and tolerates a certain number of leaked input bits. We call the resulting circuit a completely reusable garbled circuit (CRGC).
\end{enumerate}

Our CRGC scheme essentially transforms a boolean circuit $C$ that computes a functionality $f(a,b)$ into a boolean circuit $C'$ and obfuscated input $a'$ such that $C'(a',b) = C(a,b)$ for a specific input $a$ and any arbitrary input $b$. Evaluating $C'$ is as efficient as evaluating $C$. Given $C'$, $C$, and $a'$ it is difficult to infer input bits of $a$ even with repeated evaluations. Our PRGC scheme divides a CRGC $C'$ into reusable sub-circuits (sections) and non-reusable sections. Reusable sections do not leak inputs of $a$. Non-reusable sections contain gates that may leak input bits of $a$. Thus, $A$ and $B$ engage in a Yao's Garbled Circuit protocol for each non-reusable section in $C'$ for each repeated evaluation. As a result, our PRGC scheme guarantees the same level of input privacy as Yao's Garbled Circuit protocol.

Our framework compiles any user-defined C++ program and a set of inputs into a CRGC and a set of encoded inputs. $A$ can send these compressed over the network to $B$. $B$ can use our implementation to evaluate the CRGC with an arbitrary number of inputs. We tested several programs such as linear search, set intersection, and data analysis but also elementary operations such as addition and multiplication. Our benchmarks show that an Amazon M5ZN instance can construct CRGCs at approx. 80 million gates per second and evaluate them at approx. 350 million gates per second. The construction is only necessary once per input of party A. EMP SH2PC \cite{emp} and TinyGarble2 \cite{tinygarble2} can evaluate the same programs at up to 55 million gates per second but without leaking any input bits to $B$.

%% file: 02_OurApproach.tex
\section{\uppercase{Our Approach}}
\label{section:approach}

In this section, we show how $A$ can construct a CRGC and a PRGC. Any boolean circuit $C$ and generator input $a$ can be converted into an RGC $C'$ and an obfuscated input $a'$ using three different kinds of obfuscation techniques: Bit Flipping, obfuscating fixed gates, and obfuscating intermediary gates. After applying our obfuscation techniques, $C'$ is tied to a single $a'$, meaning that $\exists a \forall b: C'(a',b) = C(a,b)$ but for inputs not equal to $a'$ the output equality of $C$ and $C'$ is not ensured.    

We call $XNOR$ and $XOR$ gates \textit{balanced gates}, and all other gates \textit{imbalanced gates}. We refer to a gate as a \textit{passive gate} if modifying its truth table does not alter the circuit's output. A gate provides \textit{indistinguishability obfuscation} if $B$ has an advantage of $0$ to distinguish between a gate's truth table entry resulting from a generator's input of $0$ and $1$. In our PRGC protocol, only truth tables of gates that provide indistinguishability obfuscation and final output gates are contained in the reusable section. $A$ uses Yao's Garbled Circuit protocol to ensure that the remaining gates also do not leak any input bits of $a$. In our CRGC protocol, $A$ instead also sends these remaining gates to $B$ without additional obfuscation, thus tolerating a predictable number of leaked input bits.

\begin{figure*}

\centering

\begin{subfigure}{.5\textwidth}
  \centering
      \includegraphics [width=0.15\textwidth] {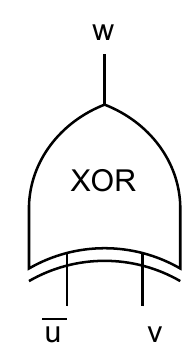}
    \rule{0cm}{0cm}
\qquad
\begin{tabular}[b]{|l|l|l|}
\hline
$\overline{u}$ & v & w  \\ \hline
0 & 0 & 1 \\ \hline
0 & 1 & 0 \\ \hline
1 & 0 & 0 \\ \hline
1 & 1 & 1 \\ \hline

\end{tabular}
  \caption{$XOR$ gate, left parent flipped.}
  \label{fig:XOR1}
\end{subfigure}%
\begin{subfigure}{.5\textwidth}
  \centering
        \includegraphics [width=0.15\textwidth] {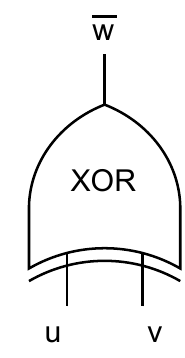}
    \rule{0cm}{0cm}
\qquad
\begin{tabular}[b]{|l|l|l|}
\hline
u & v & $\overline{w}$ \\ \hline
0 & 0 & 1 \\ \hline
0 & 1 & 0 \\ \hline
1 & 0 & 0 \\ \hline
1 & 1 & 1 \\ \hline

\end{tabular}
  \caption{$XOR$ gate, output flipped.}
  \label{fig:XOR2}
\end{subfigure}

\vskip\baselineskip

\begin{subfigure}{.5\textwidth}
  \centering
      \includegraphics [width=0.15\textwidth] {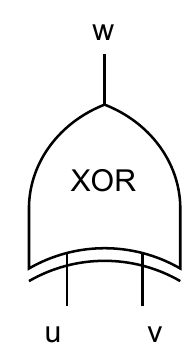}
    \rule{0cm}{0cm}
\qquad
\begin{tabular}[b]{|l|l|l|}
\hline
u & v & w \\ \hline
0 & 0 & 0 \\ \hline
0 & 1 & 1 \\ \hline
1 & 0 & 1 \\ \hline
1 & 1 & 0 \\ \hline

\end{tabular}
  \caption{$XOR$ gate, not flipped.}
  \label{fig:XOR3}
\end{subfigure}%
\begin{subfigure}{.5\textwidth}
  \centering
        \includegraphics [width=0.15\textwidth] {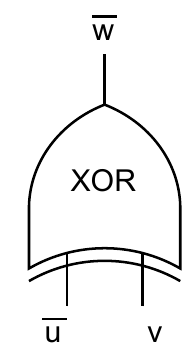}
    \rule{0cm}{0cm}
\qquad
\begin{tabular}[b]{|l|l|l|}
\hline
$\overline{u}$ & v & $\overline{w}$ \\ \hline
0 & 0 & 0 \\ \hline
0 & 1 & 1 \\ \hline
1 & 0 & 1 \\ \hline
1 & 1 & 0 \\ \hline

\end{tabular}
  \caption{$XOR$  gate, left parent \& output flipped.}
  \label{fig:XOR4}
\end{subfigure}

\caption{Flipping balanced gates yields indistinguishable truth tables.}
\label{fig:balancedFlip}
\end{figure*}

\subsection{Bit Flipping}

Bit Flipping refers to applying a one-time pad $r$ over the input bits of $a$ and all wires in the circuit $C$ to obtain $C'$ and $a'$. Only inputs from $B$ and final output wires do not get flipped. Whenever, a wire $w$ is flipped by $r$, $A$ needs to modify the truth table of $w$'s child gates to recover the integrity of $C'$. For instance, if the left input wire of a gate $g$ with functionality $f(u,v)$ is flipped, $A$ can modify $g's$ truth table to $f(\neg u,v)$ to ensure that $C'(a',b)$ = $C(a,b)$.

Since an RGC should be dependent on one fixed $a$ with a bitlength $l$, truth table entries that contain $\neg{a_i}$ ($i<l$) can be modified arbitrarily while maintaining the integrity of $C'$. Algorithm \ref{alg:bitFlipping} realizes Bit Flipping. Since a one-time pad is only secure for a single input, $A$ has to construct a new RGC if $a$ changes. However, $B$ can use $C'$ and $a'$ for multiple of its own inputs $b$. With Bit Flipping, all balanced gates ($XOR, ~XNOR$) in $C'$ achieve indistinguishability obfuscation. 

\begin{algorithm}
	\caption{Bit Flipping}
	\begin{algorithmic}[1]
        
	    \For {each generator input $a[i]$}
	       \State $a'[i] \leftarrow generateRandomBit()$
	       \State $flipped[i] \leftarrow a'[i] == a[i]$
	    \EndFor

		\For {each gate $g$}

		\State  $recoverIntegrity(g)$
		
		\State  $flipped[g] \leftarrow generateRandomBit()$
		\If {$flipped[g] == true ~\&~ g \not \in$ Output }
		\State $flipTruthTable(g)$
		\EndIf
		\EndFor

	\end{algorithmic}
	\label{alg:bitFlipping}
\end{algorithm}

\subsubsection{Examples}

Figure \ref{fig:balancedFlip} illustrates the achieved indistinguishability of randomly flipping balanced gates. Note that truth tables shown in Figure \ref{fig:XOR1} and \ref{fig:XOR2} are identical, even though their generator inputs $u$ differ. Figure \ref{fig:XOR3}, \ref{fig:XOR4} show the other two identical truth tables constructed from different inputs and flips. Since $B$ can obtain two identical truth tables from a generator's value of $0$ and $1$, it cannot infer $u$ from inspecting the truth table of a potentially flipped balanced gate.

Bit Flipping does not lead to an indistinguishability obfuscation for imbalanced gates. All four combinations of randomly flipping the generator's input $u$ and the output wire $w$ yield distinct truth tables. Thus, even though $a_i'$ is obfuscated by $r_i$, $B$ can infer $a_i$ by inspecting any imbalanced gate with functionality $f(a'_i, b_j)$. Our following two techniques also obfuscate imbalanced gates.

\subsection{Obfuscating Fixed Gates}

We define fixed gates as gates that always return the same value given the generator input $a$. For instance, an $AND$ gate that takes a generator input of $0$ is a fixed gate. The problem with an imbalanced gate on level 1 is that $B$ can immediately infer $A's$ input by observing if its output changes when changing $B's$ input bit. $A$ can effectively obfuscate those gates by flipping one of the output values in the truth table and adjusting child gates accordingly. This way, a fixed imbalanced gate at level 1 is indistinguishable from an unfixed one. When obfuscating a fixed gate, we break the gate's integrity, i.e., we might return a value of $1$ even though its correct value is $0$. The integrity of $C'$ has to be recovered to yield the correct output. Algorithm \ref{alg:HandlingFixedGates} identifies all fixed gates and ensures that modifying fixed gates maintains the correctness of $C'$. 

\begin{algorithm}
	\caption{Identify fixed gates}
	\begin{algorithmic}[1]
        
	    \For {each generator input $a[i]$}
	       \State $fixedValue[i] \leftarrow a[i]$
	       \State $isFixed[i] \leftarrow true$
	    \EndFor

		\For {each gate $g$}
		\State $l\leftarrow g.leftParent$
		\State $r\leftarrow g.rightParent$
		\State $T\leftarrow g.truthTable$
		\State $v_l \leftarrow fixedValue[l]$
		\State $v_r \leftarrow fixedValue[r]$

		\If {$isFixed[l]$ \& $isFixed[r]$}
		\State $fixedValue[g] \leftarrow T[v_l][v_r]$
		\State $isFixed[g] \leftarrow true$
		\Else  
		\If {$isFixed[l] ~\&~ T[v_l][0] == T[v_l][1]$}
				\State $fixedValue[g] \leftarrow T[v_l][0]$
		\State $isFixed[g] \leftarrow true$

		 \EndIf

		\If {$isFixed[r] ~\&~  T[0][v_r] ==  T[1][v_r]$}
				\State $fixedValue[g] \leftarrow T[0][v_r]$
		\State $isFixed[g] \leftarrow true$
		
		\EndIf
		\If {$! isFixed[g]$} 
		    \State $recoverIntegrity(g)$
		 \EndIf
		
		\EndIf

		\EndFor

	\end{algorithmic} 
	\label{alg:HandlingFixedGates}
\end{algorithm}

\subsubsection{Examples - Obfuscating Fixed Gates}

\begin{figure*}[ht]

\centering
\begin{subfigure}{.5\textwidth}
  \centering
      \includegraphics [width=0.18\textwidth] {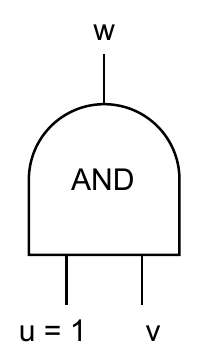}
    \rule{0cm}{0cm}
\qquad
\begin{tabular}[b]{|l|l|l|}
\hline
u & v & w  \\ \hline
0 & 0 & 1 $^\dag$ \\ \hline
0 & 1 & 0 $^\dag$ \\ \hline
1 & 0 & 0  \\ \hline
1 & 1 & 1 \\ \hline

\end{tabular}
  \caption{Unfixed $AND$ gate, obfuscated.}
  \label{fig:unfixedAND}
\end{subfigure}%
\begin{subfigure}{.5\textwidth}
  \centering
        \includegraphics [width=0.18\textwidth] {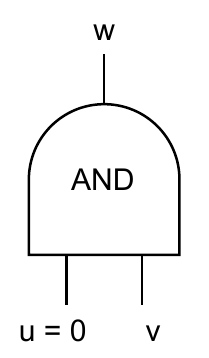}
    \rule{0cm}{0cm}
\qquad
\begin{tabular}[b]{|l|l|l|}
\hline
u & v & w \\ \hline
0 & 0 & 1 $^\star$  \\ \hline
0 & 1 & 0 \\ \hline
1 & 0 & 0 $^\dag$ \\ \hline
1 & 1 & 1 $^\dag$ \\ \hline

\end{tabular}
  \caption{Fixed $AND$ gate, obfuscated.}
  \label{fig:fixedAND}
\end{subfigure}

\caption{Obfuscated imbalanced gates on level 1.}

\rule{0in}{1.2em}$^\dag$\scriptsize These entries do not depend on the generator's input and can be re-assigned arbitrarily.\\
\rule{0in}{1.2em}$^\star$\scriptsize One relevant entry in a fixed gate gets flipped.\\
\label{fig:level1Obfuscation}

\end{figure*}

Figure \ref{fig:level1Obfuscation} illustrates the following examples. Consider an $AND$ gate $g$ at level 1 in $C$ that depends on one input $u$ of $A$ and one input $v$ of $B$. Suppose $A$'s input is $1$ (Figure \ref{fig:unfixedAND}). In this case, the relevant output entries for $g$ are $1|0|0$ and $1|1|1$ (left input$|$right input$|$output). $A$ can modify the other two entries arbitrarily as they depend on a different generator input. Thus, $A$ can obfuscate $g$ to an $XNOR$ gate by assigning the unused truth table entries to $0|0|1$ and $0|1|0$.

\begin{figure*}[ht]

\centering

\begin{subfigure}{.5\textwidth}
  \centering
      \includegraphics [width=0.18\textwidth] {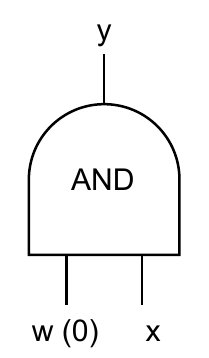}
    \rule{0cm}{0cm}
\qquad
\begin{tabular}[b]{|l|l|l|}
\hline
w & x & y  \\ \hline
0 & 0 & 0  \\ \hline
0 & 1 & 0  \\ \hline
1 & 0 & 0  \\ \hline
1 & 1 & 0 $^\dag$\\ \hline

\end{tabular}
  \caption{Child $AND$ gate.$^\star$}
  \label{fig:childFixedAND}
\end{subfigure}%
\begin{subfigure}{.5\textwidth}
  \centering
        \includegraphics [width=0.15\textwidth] {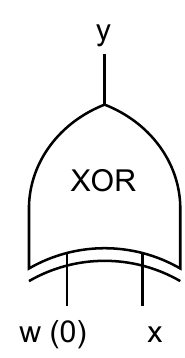}
    \rule{0cm}{0cm}
\qquad
\begin{tabular}[b]{|l|l|l|}
\hline
w& x& y\\ \hline
0 & 0 & 0 \\ \hline
0 & 1 & 1 \\ \hline
1 & 0 & 0 $^\dag$ \\ \hline
1 & 1 & 1 $^\dag$ \\ \hline

\end{tabular}
  \caption{Child $XOR$ gate.}
  \label{fig:childUnfixedXOR}
\end{subfigure}

\caption{Gates with a fixed left parent.}

\rule{0in}{1.2em}$^\dag$\scriptsize Left parent's true value is 0. Truth table gets recovered accordingly.\\
\rule{0in}{1.2em}$^\star$\scriptsize Adjusting the truth table transformed the child gate into a fixed gate. This gate gets obfuscated in the next iteration of algorithm \ref{alg:HandlingFixedGates}.\\
\label{fig:ChildGatesOfFixed}

\end{figure*}

Suppose $A$'s input is $0$ (Figure \ref{fig:fixedAND}). In this case, $g$ is a fixed gate since the two relevant entries in its truth table $0|0|0$ and $0|1|0$ both return a $0$ independent of $B$'s input. The fixed output immediately reveals $u$ to $B$ if it knows the gate type. $A$ obfuscates a fixed gate by choosing one of these entries at random and flipping its output wire. For instance, $A$ can change the entry $0|0|0$ to $0|0|1$. This way, we again created a truth table indistinguishable from $XNOR$. We showed before that $A$ can apply Bit Flipping to a balanced gate like $XNOR$ to achieve indistinguishability obfuscation. Again, truth tables shown in Figure \ref{fig:unfixedAND} and \ref{fig:fixedAND} are identical, even though their generator inputs $u$ differ.

\subsubsection{Examples - Modifying Child Gates}

Figure \ref{fig:childFixedAND} shows an $AND$ gate $g$ with a fixed obfuscated $AND$ gate as its left parent. $A$ can recover $g$'s integrity by changing the entry $1|1|1$ to $1|1|0$. $A$ just transformed $g$ into a fixed gate that always returns $0$. Thus, $A$ can apply our obfuscation technique to this gate as well. Figure \ref{fig:childUnfixedXOR} shows an $XOR$ gate with a fixed obfuscated $AND$ gate as its left parent. $A$ can recover its integrity by changing the entries $1|0|1$ and $1|1|0$ to $1|0|0$ and $1|1|1$. Notice that the resulting truth table is not a balanced gate. Thus, it does not provide indistinguishability obfuscation.

\subsection{Obfuscating Intermediary Gates}

Some gates $g_i$ do not affect the circuit's output as all paths from $g_i$ to a final output gate $g_o$ include at least one fixed gate $g_f$. We call these gates between the first level of the circuit and the dependant fixed gates \textit{intermediary gates}.

 If $A$ modifies an intermediary gate $g_i$, each fixed gate's output wire may change its value due to changing the truth table of a gate it depends on. However, we know that changing an obfuscated fixed gate's value does not change the final output of $C'$. By definition we also know that no final output gate $g_o$ directly depends on $g_i$ without a fixed gate $g_f$ between $g_i$ and $g_o$. Thus, arbitrary modifications of intermediary gates do not break the integrity of $C'$. Due to this property, fixed and intermediary gates are passive gates. $A$ can modify each passive gate's truth table to be indistinguishable from its active version. At the end of Algorithm \ref{alg:IdentifyingIntermediaryGates}, all gates where $obfuscatable[g]$ has not been set to $false$ are passive gates. Our protocol re-generates each gate on level 1 to a random balanced gate and all other passive and balanced gates to provide indistinguishability obfuscation.

\begin{algorithm}
	\caption{Identify passive gates}
	\begin{algorithmic}[1]

		\For {each final output gate $g_o$}
        \State $queue.push(g_o)$ \Comment{output gates are non-intermediary}
        \While {! queue.empty()}
        \State $g \leftarrow queue.pop()$
        \State $obfuscatable[g] \leftarrow false$ 
        \For {each parent $p$ of $g$}
        \If{$! isFixed[$p$] ~\&~ ! pushed[p]$}
        \State $pushed[p] \leftarrow true$ 
        \State $queue.push(p)$ 
        \EndIf \Comment{non-intermediary gates get pushed}
        \EndFor
        \EndWhile
		\EndFor

	\end{algorithmic} 
	\label{alg:IdentifyingIntermediaryGates}
\end{algorithm}

\subsubsection{Example}

\begin{figure}
    \centering
    \includegraphics [width=0.48\textwidth] {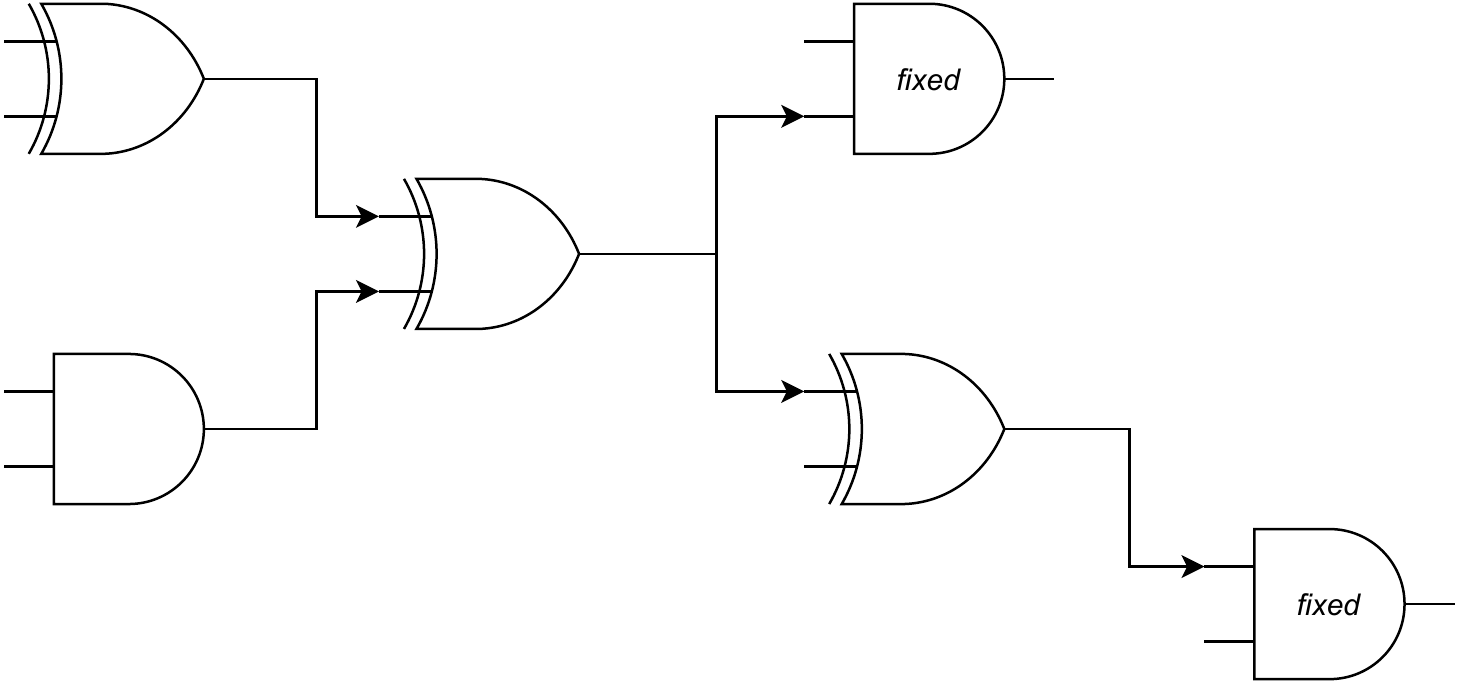}
    \caption{Section of a circuit containing four intermediary gates.$^\star$}
    \rule{0in}{1.2em}$^\star$\scriptsize Properties hold for every gate type.\\
    \label{fig:IntermediaryGate}
\end{figure}

Figure \ref{fig:IntermediaryGate} illustrates a section of a circuit with two fixed gates. Note that all paths from the unfixed gates in the section end up as an input wire of a fixed gate. Thus, all four unfixed gates in this section are intermediary gates. Modifying their truth tables may change the output of one of the fixed gates. However, this modification will not affect the output of the circuit.

\subsection{Constructing a CRGC}

The whole process of constructing a CRGC can be summarized as follows: 
\begin{enumerate}
    \item Obtain $C'$ and $a'$ by applying Bit Flipping to $a$ and all gates in $C$ using algorithm \ref{alg:bitFlipping}.
    \item Identify all fixed gates and modify all children of fixed gates in $C'$ using algorithm \ref{alg:HandlingFixedGates}.
    \item Identify all intermediary gates in $C'$ using algorithm \ref{alg:IdentifyingIntermediaryGates}.
    \item Obfuscate all imbalanced, fixed, and intermediary gates on level 1 randomly to $XOR/XNOR$. Obfuscate all passive gates in $C'$
    beyond level 1 to achieve indistinguishability obfuscation.
\end{enumerate}

After these steps, an evaluator that receives $C'$ and $a'$ can evaluate $C'$ any number of times with varying inputs $b$. For all gates $g_n$ that do not provide indistinguishability obfuscation, $B$ has an advantage $> 0$ to infer the true wire labels of $g_n$'s parents. Thus, some of these gates may leak input bits of $A$. $A$ can predict the leakage of a CRGC before constructing it and decide whether to send the CRGC that leaks some input bits or to construct a PRGC instead. PRGCs do not leak generator inputs. In the appendix, we show how $A$ can predict input leakage of a constructed CRGC and achieve n-party computations from our CRGC scheme. We also give a step-by-step example of applying our obfuscation techniques to a circuit.

\subsection{Constructing a PRGC}

After applying the three described obfuscation techniques, there is a subset of gates left in $C'$ that do not provide indistinguishability obfuscation, i.e. $B$ might be able to infer input bits of $a$ when inspecting those gates. Our PRGC scheme prevents $B$ from inferring inputs when inspecting these gates by introducing \textit{non-reusable sections}.

\subsubsection{Non-reusable Sections}

Each gate that does not provide indistinguishability obfuscation has to be contained in a non-reusable section. At the start of a non-reusable section $s$ of $C'$, $A$ and $B$ engage in Oblivious Transfer (OT) for each input wire on the first level of $s$ to let $B$ obtain keys to be used in a Yao's Garbled Circuit protocol. A non-reusable section ends if each final output gate of the non-reusable section provides indistinguishability obfuscation. $A$ needs to apply a new Bit Flipping to each of these output gates to hinder $B$ from inferring inputs by evaluating the circuit multiple times. With this approach, $A$ and $B$ have to engage in a Yao's Garbled Circuit protocol for each non-reusable section. With the output bits obtained from Yao's Garbled Circuit protocol, $B$ continues evaluating the circuit.

By "refreshing" Bit Flipping at the end of a non-reusable section, all balanced gates again provide indistinguishability obfuscation. To ensure correctness, both parties need to engage in OT for each final output gate of $C'$ to let $A$ reverse Bit Flipping applied in the non-reusable sections. Note that a simpler protocol could consist of evaluating each gate that does not provide indistinguishability obfuscation by an OT with a bit flipped result. However, if a non-reusable section spans multiple levels in the circuit, using Yao's Garbled Circuit protocol is more efficient.

\begin{figure}
    \centering
    \includegraphics [width=0.43\textwidth] {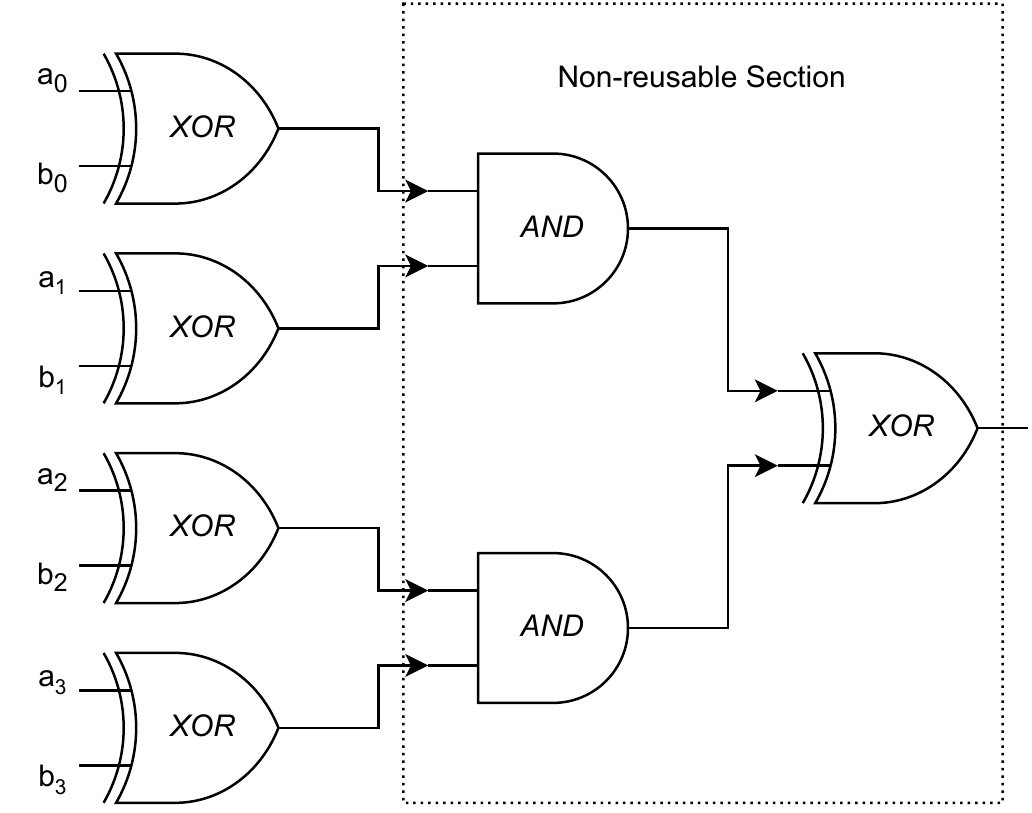}
    \caption{A circuit containing a non-reusable section.}
    \label{fig:prgc}
\end{figure}

Figure \ref{fig:prgc} illustrates a circuit with a non-reusable section. Inputs $a_i$ mark $A$'s inputs, inputs $b_i$ mark $B$'s inputs. Observe that both $AND$ gates cannot be fixed gates and reveal $A$'s input even if obfuscated by our techniques. The final $XOR$ gate marks the end of the non-reusable section by providing indistinguishability obfuscation. $A$ only sends the gates in the reusable section (first level) to $B$. By assigning the two $AND$ gates and the final $XOR$ gate to a non-reusable section, $B$ has to stop evaluating the PRGC after the first level. For each input wire of each $AND$ gate, it has to receive an input key via OT and obtain a Yao's Garbled Circuit from $A$ containing the remaining three gates. For each repeated evaluation of the circuit with different inputs $b$, it can reuse the gates on the first level of the circuit.

PRGCs provide input privacy without leakage. A security proof of PRGCs can be found in the appendix. There, we also cover how to achieve indistinguishability obfuscation for passive gates in the reusable section. In high latency environments, it might be favorable to split $C'$ into only one reusable and one non-reusable section. This way, a PRGC can be evaluated in constant communication rounds where one batch of OTs is processed in parallel.  

%% file: 03_Benchmarks.tex

\section{\uppercase{Benchmarks}}

\label{section:benchmarks}

Our open-source library is available \href{https://github.com/chart21/CRGC}{on GitHub}. With our library, $A$ can construct a CRGC from any user-defined C++ function, compress it, predict leaked input bits, and send it over the network to $B$. $B$ can evaluate the CRGC and store it on its hard drive for future use. For compiling a C++ function to a boolean circuit, we mainly rely on modules provided by EMP. 

We tested our implementation on two AWS M5ZN metal instances with 24 cores, 48 threads, and 192GB of RAM connected via 100 Gbit/s network connections to the internet in a WAN setting. $A$ can construct a CRGC at a speed of up to 85 million gates per second, perform leakage prediction with up to 115 million gates per second and evaluate a circuit with up to 395 million gates per second. For comparison, we also implemented our test programs with EMP SH2PC and TinyGarble2. Since EMP SH2PC and TinyGarble2 implement a regular garbled circuit protocol, an evaluation must always be performed together with circuit construction. In all tests, EMP performs better than TinyGarble2 and achieves a speed of up to 55 million gates per second. Thus, after only a few evaluations, our CRGC library outperforms both libraries. All CRGCs we constructed leak at most two input bits to $B$. We use the \href{https://github.com/powturbo/TurboPFor-Integer-Compression}{Turbo Pfor} integer compression algorithm \cite{pfor} before sending a CRGC over a network or storing it locally. As a result, a CRGC is approx. 75\% smaller in file size than the original uncompressed boolean circuit. 

\subsection{Basic Circuits}

Table \ref{tab:BasicCircuits} shows the results of applying our protocol to elementary circuits. $|C|$ shows the number of gates in a circuit. $Inputs ~ leaked$ refers to the number of generator inputs a CRGC leaks. All tested basic circuits can be evaluated in under 1ms by our library.

\begin{table}
	\centering
	\caption{Evaluation time for basic circuits.}
	\label{tab:BasicCircuits}
	\begin{tabular}{llllll}
		\toprule
		Circuit       & $|C|$   & Inputs &  Evaluation  \\ 
		       &    & leaked & time ($\mu$s)  \\
		
		\midrule
    64-bit Adder & 376    & 1/64  & 2 \\
    64-bit Subtract & 439    & 1/64  & 2 \\
    64-bit Multiplier & 13675 & 2/64  & 36 \\
    AES-256(k,m) & 50666 &  0/256 & 94 \\
    SHA256 & 135073 & 0/512 & 205 \\
    SHA512 & 349617 & 0/1024 & 551 \\ \bottomrule
	\end{tabular} 
\end{table}%

\subsection{Large Circuits}

We also tested more complex circuits that implement three real-world use cases: Finding an element in an unsorted list (query), identifying the maximum element in a specific coordinate range of a 2D array, and finding the intersect of two datasets. The resulting circuits have up to 1.9 billion gates and leak at most one input bit to $B$. These larger circuits demonstrate that our library is practical for real problems.

Our programs may serve as references for other functionalities. Table \ref{tab:elementsPerSeconds} shows that our library can process a dataset containing millions of entries in just a few seconds for simple functionalities. For complex functions such as the demonstrated set intersection, it can process a few thousand elements per second. These results may serve as a rough estimation of whether CRGCs can cope with a certain problem size.

\begin{figure*}[ht]
    \centering
    \includegraphics[width=0.8\textwidth]{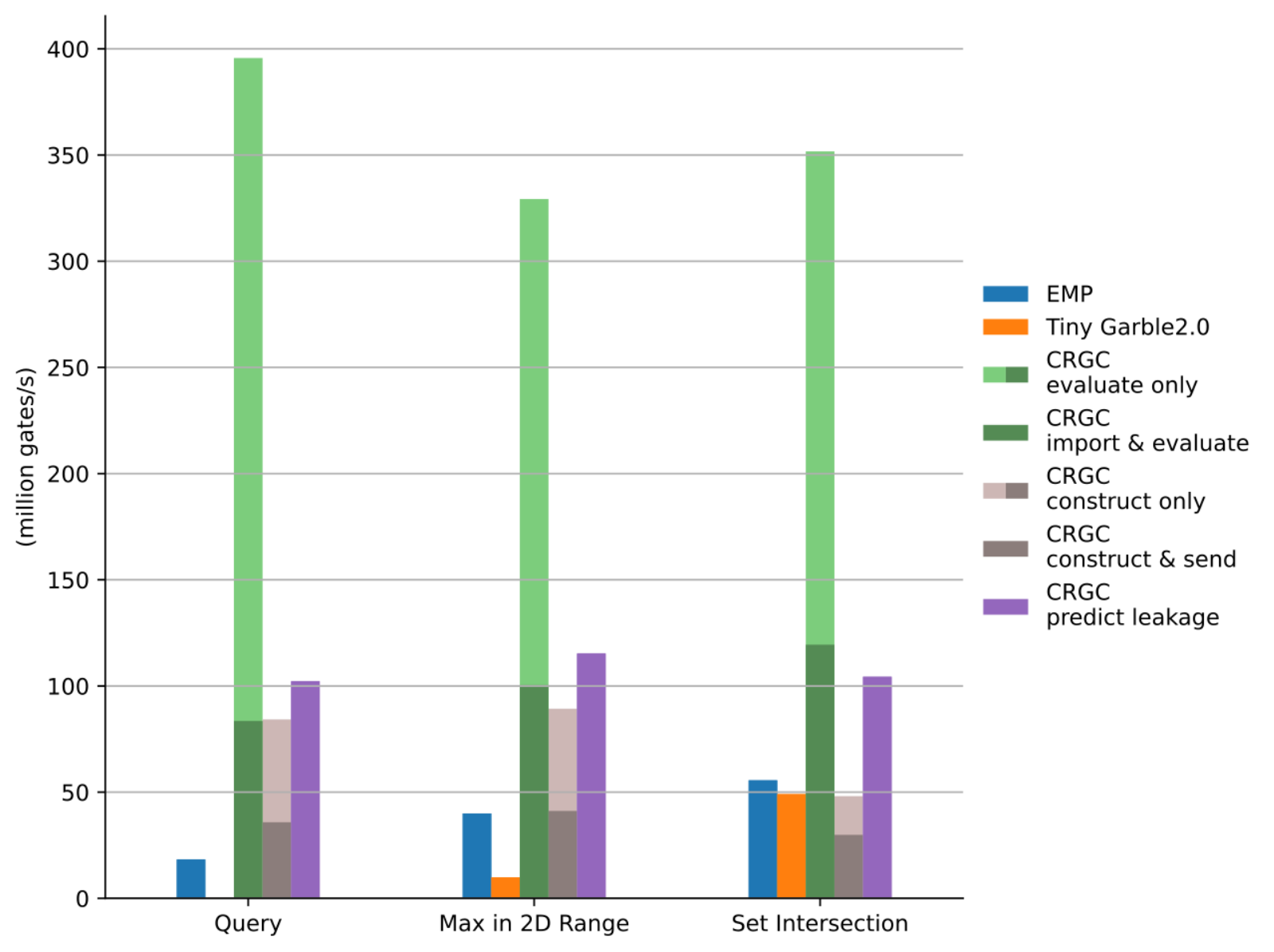}
    \caption{Performance Comparison of EMP SH2PC, TinyGarble2, and CRGC Components.}
    \label{fig:PerfComparison}
 
\end{figure*}

\begin{table*}[ht]
	\centering
	\caption{Evaluation time for large circuits.}
	\label{tab:elementsPerSeconds}
	\begin{tabular}{llllll}
	\toprule
	\multirow{2}{*}{Functionality} &  \multirow{2}{*}{$|C|$}  & \multirow{2}{*}{Elements}      & \multicolumn{3}{c}{Evaluation time (Elements/s)} \\ 
	
	& & & CRGC   & EMP SH2PC & TinyGarble2  \\ \midrule
    Query & 9,100,000 & 140,000 & 6,086,956   & 281,690   & 1,924 \\
    Max in 2D Range & 123,815,518 & 148,996 & 396,265   & 48,094   & 11,860 \\
    Set Intersection & 1,910,780,159 & 40,000 & 7,362   & 1,163 & 1,027 \\
  \bottomrule
	\end{tabular} 
\end{table*}%

Figure \ref{fig:PerfComparison} shows the results of benchmarking our framework against EMP SH2PC and TinyGarble2. Recall that $A$ has to perform leakage prediction only once per circuit, independent of its inputs. It has to generate a CRGC once per unique generator input $a$. $B$ has to evaluate a CRGC once per changing evaluator input $b$. Thus, we measured all three tasks independently. 

The bars with different shades of colors show additional costs that might occur along with a CRGC component: After $A$ constructs a CRGC, it needs to send it to $B$ over the network (brown bar). If $B$ does not store the circuit in memory after evaluating it, it needs to import the circuit from the hard drive again before performing an additional evaluation (green bar).

Our benchmark shows that our library can consistently evaluate different circuits at approx. 350 million gates per second. Constructing a CRGC and sending it to $B$ is sometimes slower than performing a regular garbled circuit protocol with EMP once. However, evaluating a CRGC is 5-20 times faster than performing a garbled circuit protocol using EMP. In all tests, using our library compared to EMP and TinyGarble2 pays off after less than three evaluations.

Note that beyond our implementation, a generator can always construct a PRGC with only one reusable and non-reusable section. Since evaluating a CRGC is faster than evaluating a regular garbled circuit, it follows that for most circuits $C$, we can construct a PRGC that can be evaluated faster than performing Yao's Garbled Circuit protocol with $C$.

%% file: 04_RelatedWork.tex
\section{\uppercase{Related Work}}

\label{section:relatedWork}

Reusable garbled circuits have been gaining popularity in the Secure Multiparty Computation community during the last decade. \cite{saleem2018recent} summarizes and discusses recent advancements in garbled circuits. The authors state that one important future step is constructing a reusable garbled circuit scheme with low computational complexity. In contrast to our approach, existing proposals tried to build CRGCs without leakage that do not scale well. To achieve practicability, we propose a trade-off between the extent of reusability and performance (PRGC), or between security and performance (CRGC).

\cite{reusable-goldwasser} proposed the first reusable garbled circuit scheme that is based on functional encryption. Functional encryption allows a user to generate secret keys that enable a key holder to learn a specific function output of encrypted data but learn nothing about the data \cite{functional_definition}. However, their scheme relies on fully homomorphic encryption and other computationally expensive techniques to achieve functional encryption. Since then, there have been optimizations to the computational complexity of reusable garbled circuits that also rely on fully homomorphic- or attribute-based encryption \cite{boneh_reusbale}.

As both prior mentioned schemes for reusable garbled circuits combine multiple complex cryptographic primitives, it is difficult to assess their efficiency. According to \cite{wangEfficientPrivateResuable2} both solutions are not practical. Thus, \cite{wangEfficientPrivateResuable2} constructed a reusable garbled circuit scheme with a trade-off between security and privacy. Their solution does not contain any benchmarks or implementations.
\cite{gorbunov_garbled_reusable} proposed a step towards reusable garbled circuits by encrypting each garbled value with a seed. For each wire and each gate, a different encryption key is used. The evaluator obtains an encoded seed in the beginning to evaluate the circuit. However, their scheme does not achieve input privacy.

Due to the lack of an existing reusable garbled circuit implementation, we compare our library with the alternative of constructing a new Yao's Garbled Circuit for each evaluation with a state-of-the-art framework. Multiple libraries have been proposed that implement Yao's Garbled Circuit protocol with various optimizations such as \textit{Free XOR} \cite{FreeXOR}. Libraries that offer state-of-the-art performance and rich functionalities are TinyGarble2 \cite{tinygarble2}, Obliv-C \cite{oblivc}, ABY \cite{aby}, and EMP SH2PC \cite{emp}. Since \cite{tinygarble2} demonstrated that TinyGarble2 outperforms Obliv-C and ABY, we chose EMP and TinyGarble2 as our benchmark.

%% file: 05_Conclusion.tex
\section{\uppercase{Conclusion}}

In this work, we proposed obfuscation-based techniques for constructing completely reusable garbled circuits (CRGCs) and partially reusable garbled circuits (PRGCs). We showed that our CRGC library can evaluate constructed circuits up to 20 times faster than current state-of-the-art garbled circuit libraries. 

CRGCs come with predictable input leakage. While we were not able to not infer multiple input bits from our test circuits, certain functionalities or more sophisticated analyses may do so. In this case, the generator and evaluator can engage in our hybrid PRGC protocol to only use a CRGC for evaluating the sections of the underlying circuit that do not pose input leakage. The remaining sub-circuits can be evaluated by Yao's Garbled Circuit protocol. Future work may introduce techniques to increase the number of gates in the reusable section or find more efficient ways to construct RGCs for n-party computation.

%% file: Appendix.tex
\section*{\uppercase{Appendix}}

\subsection*{Security Proof of PRGCs}

To prove \textit{input privacy} of a one-time protocol $\pi$ against semi-honest adversaries one can use the following two simulation proofs \cite{lindell2017simulate}:
\begin{equation}\label{eq:simproof1}
\begin{split}
\{S_1(1^n,x,f_1(x,y))\}_{x,y \in \{0,1\}^*;n \in \mathbb{N}} \\
\overset{c}{\equiv} \{view^\pi_1(x,y,n)\}_{x,y \in \{0,1\}^*;n \in \mathbb{N}}
\end{split}
\end{equation}

\begin{equation}\label{eq:simproof2}
\begin{split}
\{S_2(1^n,y,f_2(x,y))\}_{x,y \in \{0,1\}^*;n \in \mathbb{N}} \\
\overset{c}{\equiv} \{view^\pi_2(x,y,n)\}_{x,y \in \{0,1\}^*;n \in \mathbb{N}}
\end{split}
\end{equation}

\noindent Simulating $A$'s view is trivial as it does not interact with $B$ when constructing reusable sections. Simulating $B$'s view is possible by constructing a PRGC with a random generator input. Using the knowledge of $f_2(x,y)$, the simulator can modify the Oblivious Transfers required to obtain the final output bits by always returning the correct output, independent of $B's$ choice bit. Thus, we only need to show if, in $B$'s view, a PRGC based on a random generator input is indistinguishable from a PRGC based on the actual generator input.


\vspace{2mm}

\noindent \textbf{Claim} \textit{A PRGC computing an arbitrary functionality $f(a,b)$ expressed by a circuit $C$ for a fixed input $a$ and an unfixed input $b$ provides input privacy}.

\vspace{2mm}
\noindent Without loss of generality, assume $C$ only consists of balanced gates ($XOR, ~XNOR$) and the following imbalanced gates: $AND, ~NAND, ~OR, ~NOR$. Assume $G$ is a PRG that can sample a uniformly random $b$ from $U=\{0,1\}$. Assume that $B$ knows every gate $g$ in $C$. 

For each value $v_w$ at wire $w$ in $C$, $A$ samples a $b_w$ from $U$ and sets $v'_w = v_w \oplus b_w$. Each wire label $v'_w$ is now obfuscated by a one time pad. $B$ receives $b_w$ only for its own input bit wires. Let $g$ be a gate in $C$ with functionality $f(v_i, v_j) = v_k, i,j,k < |w|$. Let $B$ receive only the last column of the truth table $T_g$ from $A$ that contains all four combinations for $f(v'_i, v'_j) = v'_k$. Table \ref{tab:GenTT} shows a truth table after Bit Flipping. $p$ refers to the position in the of the last column's entry in the truth table. $B$ knows up to one input wire $v_c, b_c$ ($c \in \{i,j\}$) of $g$ in advance (let $c=j$).

If all wires satisfy the following equation, then any two PRGCs based on different generator inputs follow the same distribution of wire labels and are thus indistinguishable from $B$'s perspective. 
\begin{equation}\label{eq:Infer}
PR[v_i = 0] \overset{c}{=} PR[v_i = 1]  \overset{c}{=} \frac{1}{2}. 
\end{equation}

\noindent Thus, we reduce the proof that a PRGC provides input privacy against a semi-honest evaluator to equation \ref{eq:Infer} holding for all wires under said conditions. We split up the proof into four Lemmas. Proofing Lemma 1-3 shows that the reusable sections of a PRGC provide input privacy. Proofing Lemma 4 shows that non-reusable sections of a PRGC provide input privacy. In combination, we prove that PRGCs provide input privacy for any circuit $C$. 

\begin{table*}
\caption{Flipped gate with arbitrary functionality, denoted by $\star$.}
    \centering
   \begin{tabular}[]{|l|l|l|l|}
\hline
$p$ & $v'_i$ & $v'_j$ & $v'_k$ \\ \hline
$2b_i + b_j$ & $0 \oplus b_i$ & $0 \oplus b_j$ & $[(0 \oplus b_i) \star (0 \oplus b_j)] \oplus b_k$ \\ \hline
$2b_i + 1 - b_j$ & $0 \oplus b_i$ & $1 \oplus b_j$ & $[(0 \oplus b_i) \star (1 \oplus b_j)] \oplus b_k$ \\ \hline
$2 - 2b_i + b_j$ & $1 \oplus b_i$ & $0 \oplus b_j$ & $[(1 \oplus b_i) \star (0 \oplus b_j)] \oplus b_k$ \\ \hline
$2 - 2b_i + 1 - b_j$ & $1 \oplus b_i$ & $1 \oplus b_j$ & $[(1 \oplus b_i) \star (1 \oplus b_j)] \oplus b_k$ \\ \hline

\end{tabular}
    
    \label{tab:GenTT}
\end{table*}

\vspace{5 mm}

\noindent \textbf{Lemma 1} The position $p$ of an entry $v_k$ in the truth table $T_g$ does not leak $v_i$ under the security assumptions of $G$. 

\vspace{1 mm}

\noindent \emph{Proof:} In the unmodified truth table of $g$, $B$ can infer the following from $p$:
\begin{equation}
    p \in \{0,1\} \implies v_i = 0 
\end{equation}
\begin{equation}
    p \in \{2,3\} \implies v_i = 1
\end{equation}

\noindent After Bit Flipping $B$ can infer the following from $p$:
\begin{equation}
    p \in \{0,1\} \implies v_i \oplus b_i  = 0
\end{equation}
\begin{equation}
    p \in \{2,3\} \implies v_i \oplus b_i  = 1
\end{equation}

\noindent Since $B$ does not hold $b_i$, it cannot infer $v_i$ from its position $p$ in the truth table without breaking the security assumptions of $G$.  

\vspace{5 mm}

\noindent \textbf{Lemma 2} A balanced gate $g$ in the reusable section does not leak $v_i$ under the security assumptions of $G$.

\vspace{1 mm}
\noindent \emph{Proof:} The following equation holds if $g$ is an $XOR$ gate: 

\begin{equation}
\begin{split}
v'_k &= v'_i \oplus v'_j \\
&= v_k \oplus b_k = v_i \oplus b_i \oplus v_j \oplus b_j 
\end{split}
\end{equation}
\noindent The following equation holds if $g$ is an $XNOR$ gate:

\begin{equation}
v_k \oplus b_k = \neg (v_i \oplus b_i \oplus v_j \oplus b_j) 
\end{equation}

\noindent Since B does not hold any values of $\{v_i \oplus b_i, v_k \oplus b_k\}$, $B$ cannot distinguish between the entries in $T_g$ where $v_i = 1$, $v_i = 0$, $v_k = 0$, $v_k = 1$. Thus, $B$ cannot infer $v_k$ and $v_i$ when inspecting the truth table $T_g$ of balanced gate $g$ without breaking the security assumption of $G$.

\vspace{5 mm}

\noindent \textbf{Lemma 3} An imbalanced gates $g$ in the reusable section does not leak $v_i$ under the security assumptions of $G$.

\vspace{1 mm}

\noindent \emph{Proof:} The following equation holds if $g$ is an $AND$ gate:

\begin{equation}
v_k \oplus b_k = v_i \oplus b_i \wedge v_j \oplus b_j 
\end{equation}

\noindent With a certain probability $q$, $A$ replaces $g$ by a $NOR$ gate:

\begin{equation}
v_k \oplus b_k = \neg (v_i \oplus b_i \lor v_j \oplus b_j) 
\end{equation}

\noindent Only $v_i = 1$ can lead to the unique output of an $AND$ gate ($v_k = 1$). Only $v_i = 0$ can lead to the unique output of a $NOR$ gate ($v_k = 1$). By inspecting the truth table of this term $B$ finds identical values of $v'_k$ for three cases. In the other case it can infer that $v_i$ produces the unique output of $g$. However, if it cannot distinguish if $A$ replaced $g$ before Bit Flipping it cannot infer the value of $v_i$. Thus $A's$ goal is to replace $g$ with a probability $q$ such that from $B's$ perspective $Pr[g \in NOR] = Pr[g \in AND]$. 

Replacing $g$ does not maintain the integrity of $C'$. Thus, $A$ can only replace $g$ if it is a passive gate. Let $s$ be the set of possible input combinations for $a$ where $g$ is a passive gate. $A$ and $B$ can calculate $p = Pr[g \in \{$passive gates$\}] =  \frac{|s|}{2^{|a|}}$. $A$ sets the probability to replace $g$ to:

\begin{equation}
\begin{split}
pq = (1-p) + (1-q)p \\
\qquad \Leftrightarrow 2q = \frac{1-p}{p} + 1 \\
 \Leftrightarrow q =  \frac{1}{2p}
\end{split}
\end{equation}

\noindent If $q \leq 1$, $A$ replaces $g$ with probability $q$ to achieve:
\begin{equation}
Pr[g \in NOR] \overset{c}{=} Pr[g \in AND] \overset{c}{=} \frac{1}{2}. 
\end{equation}
\noindent If $q > 1$, $A$ must not add $g$ to the reusable section of $C'$. With the same procedure, A can securely obfuscate $NOR$ gates ($AND$ gates as replacement), $NAND$ gates ($OR$ gates as replacement), and $OR$ gates ($NAND$ gates as replacement). After replacing (or not replacing) $g$, A applies Bit Flipping to $g$.

By proofing Lemma 1-3, we showed that a reusable section containing only the balanced gates of $C$ and imbalanced gates that meet the conditions above satisfies equation \ref{eq:Infer}. All other gates of $C'$ are contained in a non-reusable section. 

\vspace{5 mm}

\noindent \textbf{Lemma 4} A gate $g$ in the non-reusable section does not leak $v_i$ under the security assumptions of Yao's Garbled Circuit protocol.

\vspace{1 mm}

\noindent \emph{Proof:} For each gate $g$ in the first level of a non-reusable section $s$, $B$ holds both input wires $v'_i$, $v'_j$. By engaging in two Oblivious Transfers per gate with $A$, $B$ obtains two input keys per gate. $A$ garbles the circuit $s$ according to Yao's Garbled Circuit protocol. Yao's Garbled Circuit protocol was proven to be secure before \cite{lindell2009proof}. Each final output gate of $s$ is either also a final output gate of $C$ or meets the conditions of a gate contained in the reusable section. In the former case, there is no difference to Yao's Garbled Circuit protocol. In the latter case, equation \ref{eq:Infer} holds as proven in lemma 1-3 \qedsymbol. 

If $A$ does not need to learn the output of the computation, a PRGC is secure against a malicious $B$ when utilizing a compatible OT protocol.

\subsection*{Enabling n-party Computation with CRGCs}

Our CRGC protocol can be easily extended to enable n-party computation. The following steps are neccessary for 3-PC:
\begin{enumerate}
    \item A sends $C'$ and $a'$ to party B.
    \item Party B further obfuscates $C'$ and its inputs $b$ and sends $C''$, $a'$, and $b'$ to party C.
    \item Party C can evaluate $C''$ with $a'$, $b'$, and arbitrary inputs $c$.
\end{enumerate}

If party C wants to evaluate $C''$ with different inputs of B, the parties have to repeat steps 2-3. For different inputs of party A, they have to repeat all steps. Thus, the order of parties receiving and further obfuscating a CRGC is relevant. We can generalize this observation for n-party computation. If any party wants to evaluate the circuit with a different input of a party at position $i$ in the receiving order, all parties at position $p$ with $i \leq p \leq n-1$ need to repeat obfuscation.   

\newpage

\subsection*{Predicting Leakage of a CRGC}

To predict leaked input bits of a CRGC, we have to take the evaluator's perspective when it receives $C'$ and $a'$ from $A$. By default, $B$ does not know whether a gate in $C'$ is obfuscated or flipped except for a final output gate (that is never obfuscated nor flipped). However, if it knows $C$'s exact construction, it can identify gates that are not passive or balanced with certainty. These gates do not provide indistinguishability obfuscation and may reveal input bits of $A$.

\subsubsection*{Potentially Fixed Gates}

First, we introduce the concept of \textit{potentially fixed gates}. From $B$'s perspective, any imbalanced gate on level 1 is a potentially fixed gate. As we showed before, we achieved indistinguishability obfuscation for all gates on level 1. However, in deeper levels of $C$, $B$ may identify gates that $A$ could not have obfuscated.

To identify potentially fixed gates, $B$ can use the following ruleset. It can consider all generator inputs as potentially fixed and all evaluator inputs as not potentially fixed. A balanced gate that has at least one not potentially fixed parent is not potentially fixed itself. This property holds because evaluating a balanced gate such as $XOR$ with one fixed and one unfixed bit always returns two different output bits. An imbalanced gate instead is only not potentially fixed if both parents are not potentially fixed. This property holds because evaluating an imbalanced gate such as $AND$ with at least one fixed bit may always return the same output bit. $B$ can iterate through the whole circuit with this ruleset to identify all not potentially fixed gates. Algorithm \ref{alg:pfGates} applies this ruleset to a CRGC.

\begin{algorithm}
	\caption{Identify potentially fixed gates.}
	\begin{algorithmic}[1]

\algnewcommand\algorithmicswitch{\textbf{switch}}
\algnewcommand\algorithmiccase{\textbf{case}}
\algnewcommand\algorithmicassert{\texttt{assert}}
\algnewcommand\Assert[1]{\State \algorithmicassert(#1)}%
\algdef{SE}[SWITCH]{Switch}{EndSwitch}[1]{\algorithmicswitch\ #1\ \algorithmicdo}{\algorithmicend\ \algorithmicswitch}%
\algdef{SE}[CASE]{Case}{EndCase}[1]{\algorithmiccase\ #1}{\algorithmicend\ \algorithmiccase}%
\algtext*{EndSwitch}%
\algtext*{EndCase}%

	    \For {each generator input $a'_i$}
	       \State $pf[a'_i] \leftarrow true$ \Comment{$pf \widehat{=}$ potentially fixed}
	    \EndFor
	    
	    \For {each evaluator input $b_i$}
	       \State $pf[b_i] \leftarrow false$
	    \EndFor

		\For {each gate $g$}

		\Switch {type(g)}
\Case {$type(g) \in imbalancedGates$}
  \State $pf[g] = pf[l] \lor pf[r]$
\EndCase
\Case {$type(g) \in $ \{$XOR,XNOR$\}}
  \State $pf[g] = pf[l] \wedge pf[r]$
 \EndCase
\Case {$type(g) \in $ \{\{$0,0,1,1$\},\{$1,1,0,0$\}\}}
    \State $pf[g] = pf[l]$
\EndCase
\Case {$type(g) \in $ \{\{$0,1,0,1$\},\{$1,0,1,0$\}\}}
    \State $pf[g] = pf[r]$ 
\EndCase
\Case {$Default$} \Comment{\{$0,0,0,0$\} or \{$1,1,1,1$\}}
  \State $pf[g] = true$  
\EndCase 
\EndSwitch

		\EndFor

	\end{algorithmic} 
	\label{alg:pfGates}
\end{algorithm} 

\subsubsection*{Potentially Intermediary Gates}

Recall that intermediary gates and fixed gates provide indistinguishability obfuscation. Thus, $B$ also needs to identify all \textit{potentially intermediary gates}. If it assumes all potentially fixed gates to be fixed gates, it can identify intermediary gates by algorithm \ref{alg:IdentifyingIntermediaryGates}. As the set of fixed gates is a subset of all potentially fixed gates, $B$ can identify all potentially intermediary gates this way.

\subsubsection*{Potentially Revealing Gates}

Recall that Bit Flipping provides indistinguishability obfuscation only for balanced gates. Thus, $B$ can identify the true values of both parents of a gate with certainty if the gate is not balanced and not potentially passive. However, identifying such a gate does not yield input leakage yet. Thus, we call those gates \textit{potentially revealing gates}. 

Suppose there is a potentially revealing gate on level 1. Since a potentially revealing gate, $g_{pr}$ always reveals the true value of its parents to $B$ (if it knows $C$), a potentially revealing gate on level 1 would leak its input bits. However, each revealing gate is located at a deeper level of $C'$. Therefore, its leakage does not always reveal an input bit of $A$. 

Consider a potentially revealing gate $g_{pr}$. $B$ can only infer a generator input bit $a'_i$ if there is at most one balanced gate on the path of $g_{pr}$ to $a'_i$ since it does not know whether a balanced gate's input wires are flipped. This approach is utilized by our library to predict the number of inputs leaked in a CRGC. 

However, $B$ could combine the knowledge of multiple potentially revealing gates by setting up a system of boolean equations from each input bit to each revealed value. Calculating solutions to this system of equations may require an exhaustive search and is infeasible for large circuits. Thus, we do not provide an implementation for this approach. This means, however, that our implemented leakage prediction only serves as a lower bound. We also do not exclude the possibility that there are more ways to infer input bits from potentially revealing gates. In case a CRGC leaks multiple input bits, $A$ can construct a PRGC instead. 

Alternative threat models than discussed here may assume that $B$ does not know $C$'s construction. In this case, passive gates can be re-generated completely at random instead of providing indistinguishably obfuscation. This is the default setting of our library.

\subsection*{Example - Constructing a CRGC}

Figure \ref{fig:obfExample} illustrates an exemplary section $s$ of a circuit and shows the key modifications when using our three obfuscation techniques. Figure \ref{fig:exampleSection} shows the plain circuit and its inputs. For simplicity, we use only $AND$ and $XOR$ gates to cover one type of balanced and one type of imbalanced gates. In the example, the left parent of each gate on level 1 is always $A$'s input, and the right parent is always $B$'s input. Since useful real-world circuits are too large to illustrate in an example, we assume that the shown sequence of gates is only a section of a bigger circuit.

\subsubsection*{Bit Flipping}

Figure \ref{fig:exampleBitFlipping} illustrates how each gate and input bit is modified when Bit Flipping is applied. At first, $A$ generates obfuscated inputs. The suffix $(!)$ next to a wire value indicates that the obfuscated input is the flipped version of the original input. Bit Flipping first recovers the integrity of each gate if one of its parents got flipped. Afterward, with a probability of $\frac{1}{2}$, the output wire of each gate gets flipped as well. In the figure, the four values inside each gate show the output entries of the sorted truth table after the recovery step. Two columns inside a gate indicate that the gate also got flipped afterward. In this case, the values on the left show the truth table after recovering the gate's integrity, while the values on the right show the truth table after the output wire got flipped. Notice that the $XOR$ gate with evaluator input bit $b$ ($XOR_1$) provides indistinguishability obfuscation since $B$ cannot distinguish $XOR_1$'s truth table from the one where $A's$ input is $0$ and $XOR_1$ is flipped.

\subsubsection*{Obfuscating Fixed Gates}

Figure \ref{fig:exampleObfuscationFixed} illustrates how fixed gates and their parents get modified after Bit Flipping is applied. The four most left values inside each gate show the truth table of each gate after Bit Flipping. \textit{obf/o} indicates that a gate is fixed and shows the truth table after obfuscating it. Recall that all gates on level 1 of the circuit get obfuscated to a gate indistinguishable from $XOR$/$XNOR$. \textit{L1} indicates that an unfixed imbalanced gate gets obfuscated into a balanced gate. \textit{rec/r} indicates that the child $g_c$ of a fixed gate gets modified to recover the circuit's integrity. If this modification leads to $g_c$ being fixed, it gets obfuscated afterward (indicated by \textit{o}).

Notice that after applying this obfuscation technique to all gates on level 1, each gate's truth table is indistinguishable from either $XOR$ or $XNOR$. Observe that after recovering a gate's integrity, each truth table gets modified to be independent of its obfuscated parent. After recovery, any modification to the obfuscated gate does not change the output of its children. All fixed gates get modified to provide indistinguishability obfuscation.

\subsubsection*{Obfuscating Intermediary Gates}

\begin{figure*}

\centering

\begin{subfigure}{.5\textwidth}
  \centering
      \includegraphics [width=1\textwidth] {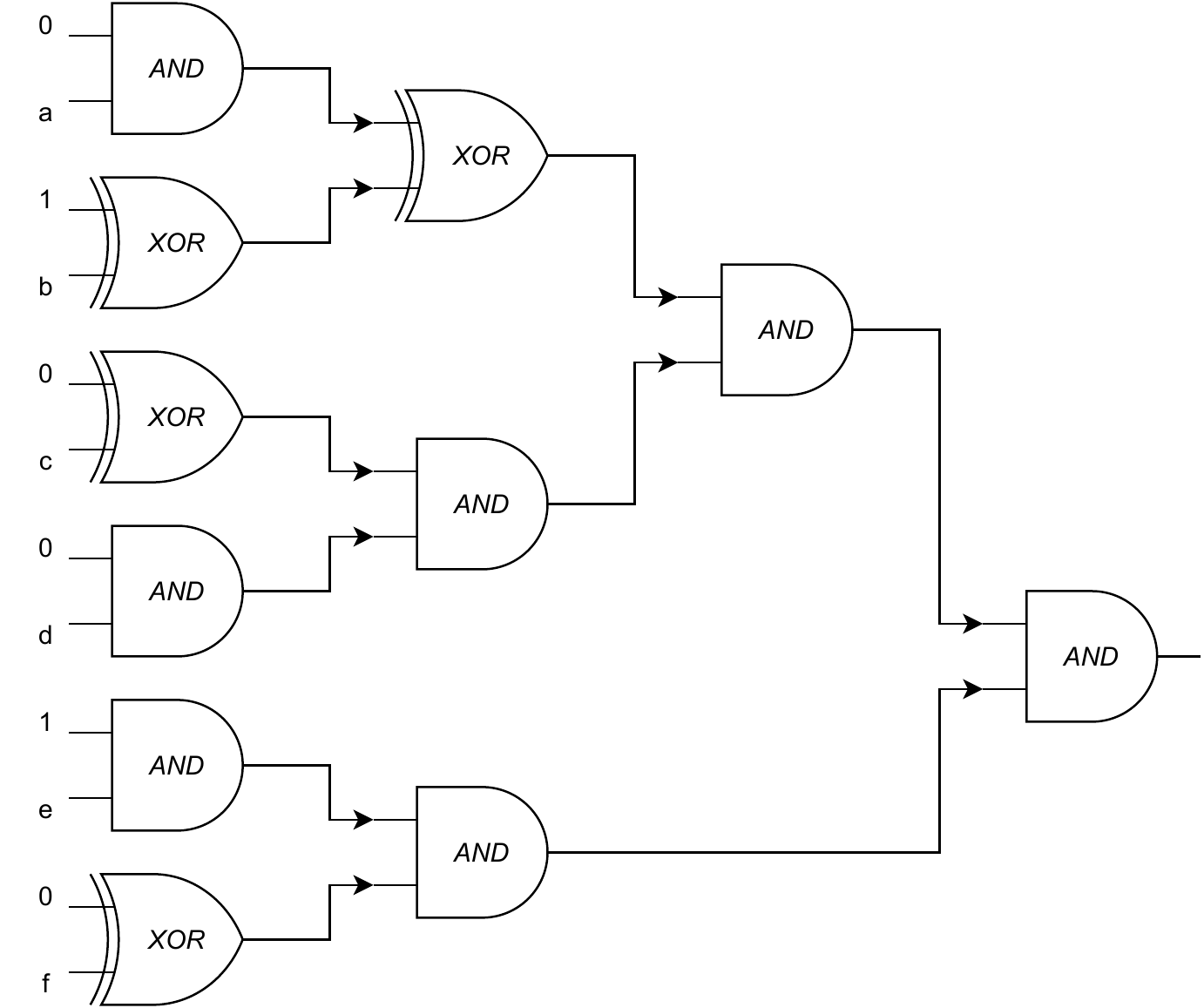}

  \caption{Section of a circuit.}
  \label{fig:exampleSection}
\end{subfigure}%
\begin{subfigure}{.5\textwidth}
  \centering
        \includegraphics [width=1\textwidth] {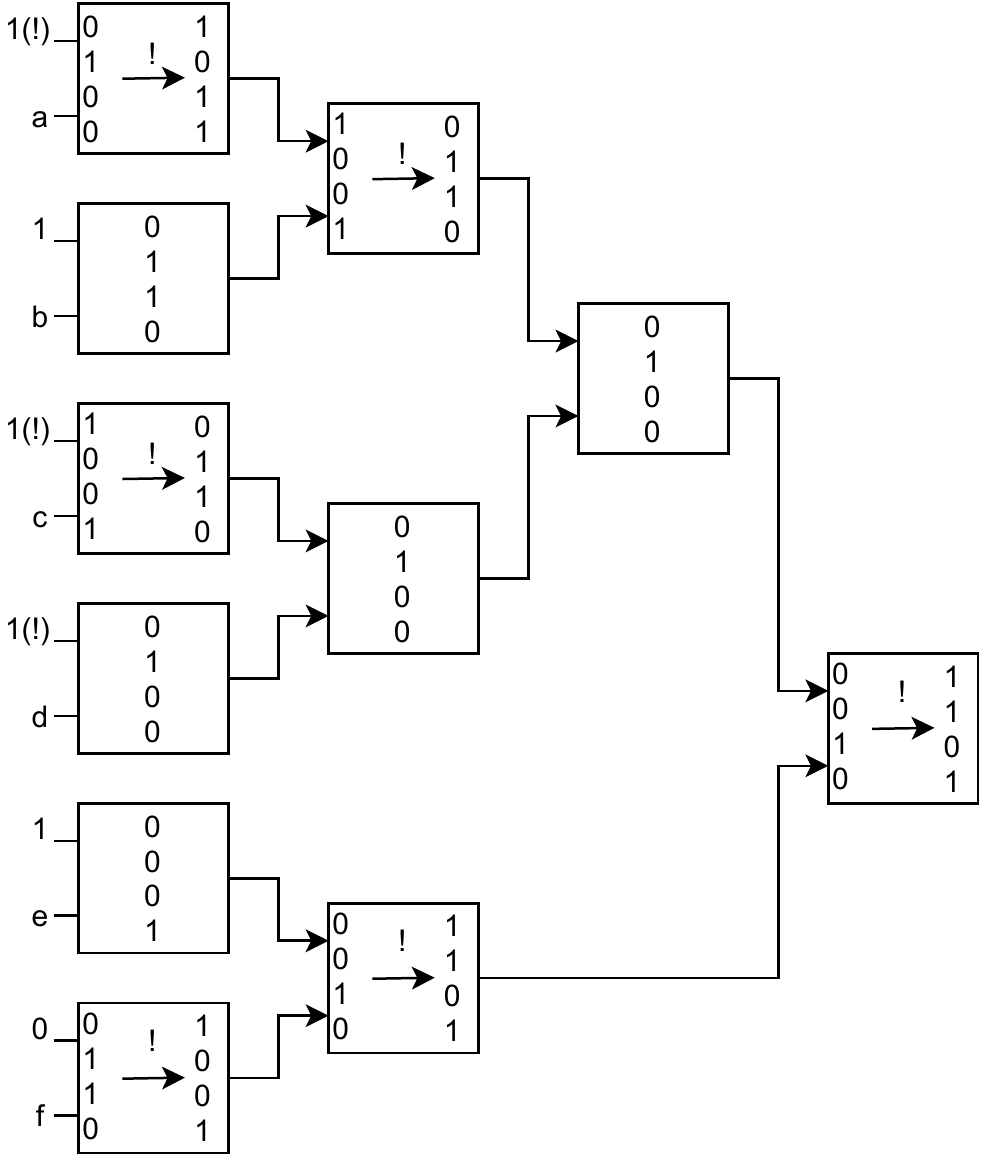}
 
  \caption{Bit Flipping.}
  \label{fig:exampleBitFlipping}
\end{subfigure}

\vskip\baselineskip

\centering

\begin{subfigure}{.5\textwidth}
  \centering
      \includegraphics [width=1\textwidth] {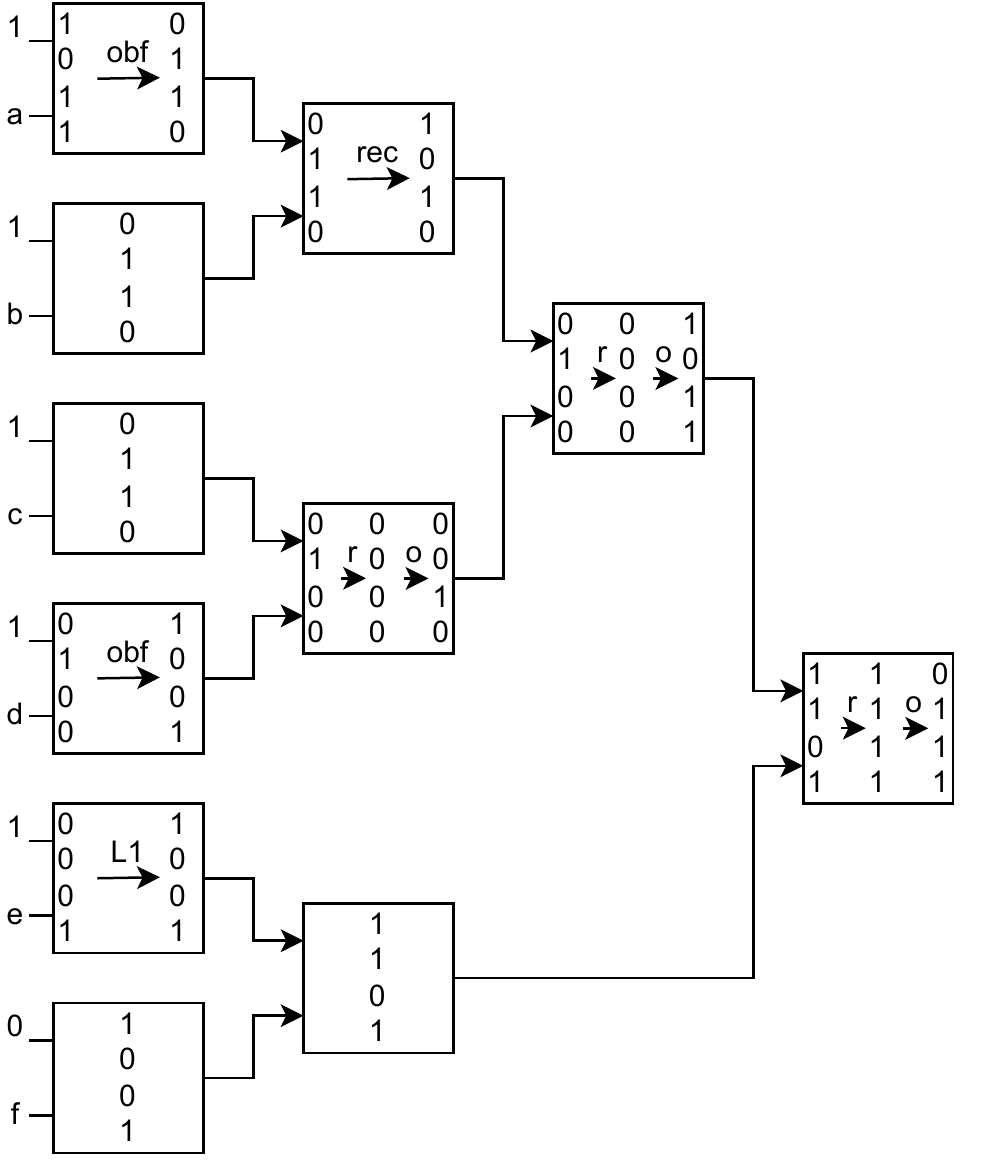}

  \caption{Obfuscating fixed gates.}
  \label{fig:exampleObfuscationFixed}
\end{subfigure}%
\begin{subfigure}{.5\textwidth}
  \centering
        \includegraphics [width=1\textwidth] {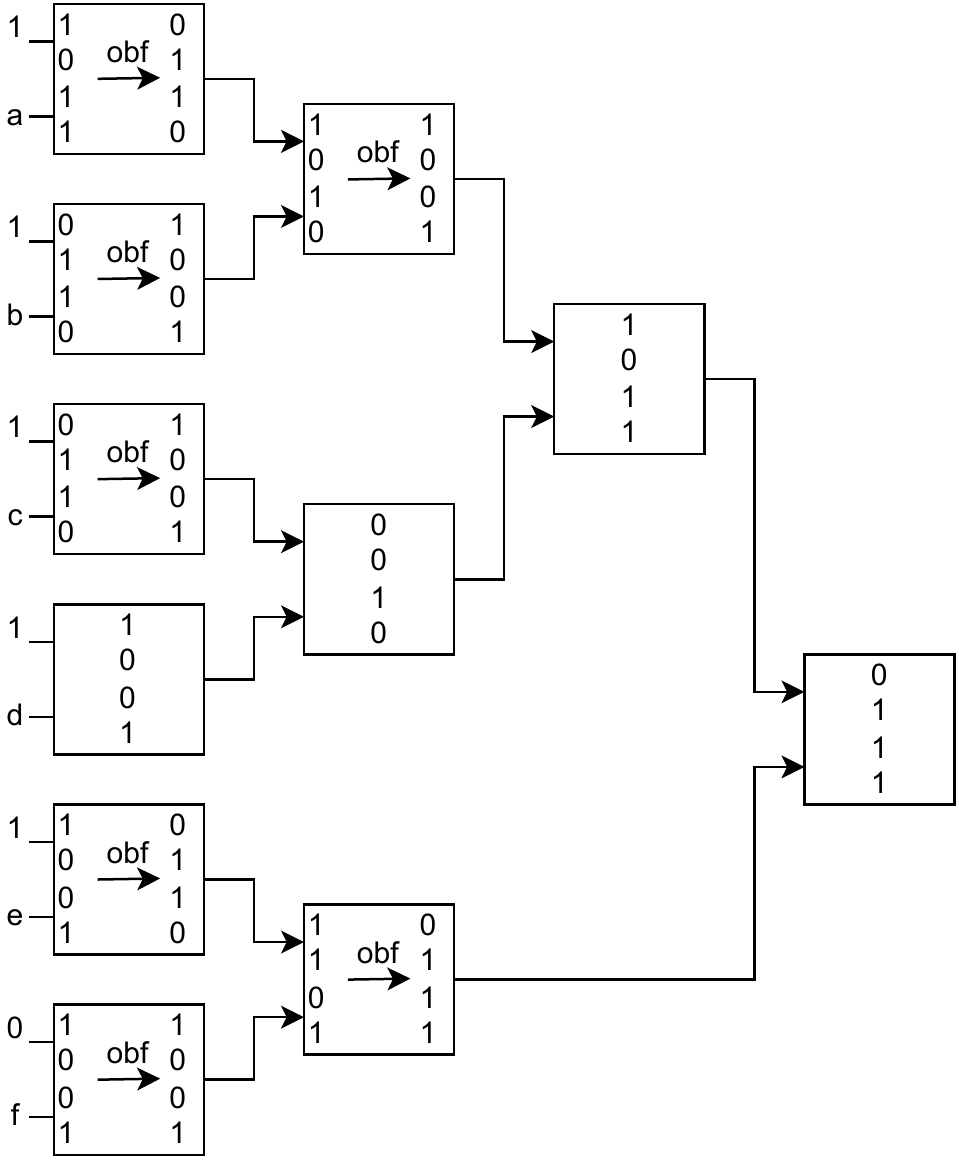}
 
  \caption{Obfuscating intermediary gates.}
  \label{fig:exampleObfuscationIntermediary}
\end{subfigure}

\caption{Constructing a CRGC.}

\label{fig:obfExample}

\end{figure*}

Figure \ref{fig:exampleObfuscationIntermediary} illustrates how intermediary gates get modified after obfuscating fixed gates. Recall that each gate $g$ where each path from $g$ to a final output gate $g_o$ contains a fixed gate $g_f$ is an intermediary gate. $A$ can modify these gates to achieve indistinguishability obfuscation without breaking the circuit's integrity. The four most left values inside each gate show the truth table of each gate after obfuscating fixed gates.\textit{obf} indicates that this intermediary gate gets obfuscated. Since the final gate of $s$ in this example is an obfuscated fixed gate, all other gates are intermediary gates. Thus, $A$ can obfuscate all gates to provide indistinguishability obfuscation. After applying our obfuscation techniques to the whole circuit, $A$ can send $C'$ and $a'$ to $B$.

\begin{figure*}
\vspace{1cm}
\centering

Codebase: \url{https://github.com/chart21/CRGC}

\end{figure*}